\documentclass[12pt]{iopart}
\usepackage{iopams}
\usepackage{epsf}

\begin{document}

\title[Electron-atom ionization near the Bethe ridge]
{Electron-atom ionization near the Bethe ridge: revision of plane
wave first-order theories}

\author{Konstantin A Kouzakov$^1$, Pavel S Vinitsky$^2$, Yuri V Popov$^2$
and Claude Dal Cappello$^3$}

\address{$^1$ Department of Nuclear Physics and Quantum Theory of Collisions,
Faculty of Physics, Moscow State University, Moscow 119992, Russia}

\address{$^2$ Nuclear Physics Institute, Moscow State University, Moscow 119992, Russia}

\address{$^3$ Universit\'e Paul Verlaine-Metz, Laboratoire de
Physique Mol\'eculaire et des Collisions, ICPMB (FR 2843),
Institut de Physique, 1 rue Arago, 57078 Metz Cedex, France}
\ead{kouzakov@srd.sinp.msu.ru}

\begin{abstract}
We consider electron impact single ionization of an atom at large
energy-momentum transfer in the nearly Bethe-ridge kinematics. For
evaluation of the ionization amplitude, a plane wave Born series
is employed. A regularization procedure is utilized in
circumventing typical divergence problems associated with the
higher-order Born terms. The regularized Born series for the
ionization amplitude is derived. On this basis, renormalized
analogs of the traditional plane wave Born and impulse
approximations are developed. These renormalized first-order
models resemble the traditional plane wave impulse approximation
with a modified Gamow factor. Numerical results using different
approximations are presented and analyzed for the case of
electron-hydrogen ionization. The present theoretical
consideration can be important for absolute measurements.
\end{abstract}

\pacs{34.80.Dp, 03.65.Nk, 34.10.+x}

\submitto{\JPB}

\maketitle

\section{Introduction}
\label{intro}
%
Single ionization by electron impact (or (e,2e) collision) at
large energy-momentum transfer in the nearly Bethe-ridge
kinematics, where the recoil ion momentum is small compared with
the transferred one, constitutes a powerful spectroscopic tool for
exploring electronic structure of atomic
systems~\cite{weigold_book,neud,watanabe05,ren05}. This (e,2e)
method is often referred to as the Electron Momentum Spectroscopy
(EMS)~\cite{weigold_book,neud}. Theoretical grounds of the EMS
exploit domination of the lowest-order term, such as given by the
plane wave Born or impulse approximation (PWBA or PWIA), when
treating the ionization amplitude perturbatively and using plane
waves for description of the incident and outgoing electron
states. Formally, the ratio of the higher-order and lowest-order
Born terms behaves as~$\propto E_0^{-1/2}$, where $E_0$ is the
incident electron energy, and hence one might expect the
lowest-order term to prevail at high values of $E_0$. However, on
the energy shell the plane wave higher-order terms are given by
divergent integrals~\cite{popov81}, which are due to so-called
Coulomb singularities of the transition
operator~\cite{latypov93,shablov99,shablov02}. Therefore, for
drawing conclusions on the validity of the PWBA model, one should
cope with the problem of divergences of the corresponding
perturbation series.

The Born series follows from the Lippmann-Schwinger equation which
has a noncompact kernel. However, in the case of short-range
potentials, the noncompactness of the kernel does not prevent from
calculating matrix elements of the Born series. For the purpose of
a mathematically correct formulation, Faddeev proposed to reduce
the Lippmann-Schwinger equation to a system of three coupled
equations with compact kernels~\cite{faddeev1,faddeev2}. This
system leads to the Born-Faddeev series where two-particle
amplitudes depending on three arguments (the relative-motion
energy and incoming/outgoing momenta) appear instead of the
corresponding potentials. The higher-order Born-Faddeev terms
contain the two-particle amplitudes which in general are
off-shell, i.e. their arguments do not necessarily obey the
energy-momentum relations for free particles. The lowest-order
Born-Faddeev term, given by a half-on-shell two-particle
amplitude, yields the usual PWIA.

In the case of Coulomb potentials Faddeev's reduction does not
offer mathematical advantages, since the kernel remains
noncompact~\cite{latypov93,faddeev2}. This is a consequence of the
fact that the plane wave states do not obey the correct asymptotic
behavior for the Coulomb breakup. Thus, the resultant Born-Faddeev
series contains divergent terms and, like the PWBA case, the
validity of the PWIA model is questionable. At the same time,
plane waves are usually a convenient and handy mathematical tool
for calculating the Born and Born-Faddeev series. And as far as
the corresponding higher-order terms diverge on the energy shell,
one must resort to a regularization procedure which removes the
artificial, unphysical problem of divergences. Further, one must
find a relation between the regularized perturbation series and
the exact ionization amplitude. Only then, a perturbative
treatment of the ionization amplitude can be developed in a
physically consistent manner.

The objective of this work is to revise the traditional PWBA and
PWIA models in the light of the fact that these first-order
theories ignore the divergence problem associated with the
higher-order contributions to the ionization amplitude. The
present theoretical analysis is based on the results
of~\cite{popov81}, where a practical recipe for regularization of
the plane wave Born series was proposed, and those of Shablov
\emph{et al}~\cite{shablov99,shablov02,faddeev2}, who established
a relation between the exact ionization amplitude and the
unphysical plane wave Lippmann-Schwinger one in the on-shell
limit. We proceed from the Born series, since, as mentioned above,
Faddeev's reduction is not of benefit in the case of Coulomb
potentials and hence may lead to methodological confusion. After
regularizing the Born series, one can formulate renormalized
analogs of the traditional PWBA and PWIA models. These analogs
resemble the traditional PWIA model with a modified Gamow factor.
As shown below, the modified Gamow factor depends on the choice of
a regularization procedure and therefore it is not uniquely
determined. This feature permits, in principle, to choose such
version of the Gamow factor that efficiently incorporates
higher-order effects ignored by the traditional first-order
models. The above points are addressed in the consideration that
follows and are illustrated with numerical results.

Section~\ref{genform} of this paper delivers a general formulation
for the (e,2e) reaction on an atom. In section~\ref{regproc},
specific Born approximations are formulated using a regularized
Born series. Section~\ref{impapprox} is devoted to the impulse
approximation theory in the context of the regularization
formalism. The numerical results for the case of electron-hydrogen
ionization are presented and discussed in section~\ref{numres} and
the conclusions are drawn in section~\ref{concl}. The atomic units
(au) $e=\hbar=m_e=1$ are used throughout unless otherwise stated.

\section{General formulation}
\label{genform}
We specify the momenta of the incident,  scattered and ejected
electrons by ${\bi k}_0$, ${\bi k}_s$ and ${\bi k}_e$,
respectively. The corresponding energies are denoted by $E_0$,
$E_s$ and $E_e$. The initial atomic and the final ionic states are
specified by their respective wavefunctions $\Phi_i^Z$,
$\Phi^{Z-1}_f$, where $Z$ designates the nuclear charge, and
energies $\varepsilon_i$, $\varepsilon_f$. The rate of the (e,2e)
reaction is characterized by the triple differential cross section
(TDCS)
\begin{equation}
\label{tdcs} \frac{\rmd^3\sigma}{\rmd\Omega_s\rmd\Omega_e\rmd
E_e}=\frac{k_ek_s}{(2\pi)^5k_0}
\left(\frac{1}{4}|{T}_s+{T}_e|^2+\frac{3}{4}|{T}_s-{T}_e|^2\right).
\end{equation}
Here the directions of the outgoing electron momenta are specified
by the solid angles $\Omega_s$ and $\Omega_e$. The amplitude
${T}_s$ (${T}_e$) corresponds to the situation where the scattered
electron has the momentum ${\bi k}_s$ (${\bi k}_e$). The so-called
capture amplitude is ignored, since we consider such kinematical
regimes where the capture of the incident electron accompanied
with ejection of two atomic electrons having the momenta ${\bi
k}_s$ and ${\bi k}_e$ is negligible. In equation \eref{tdcs} a sum
(average) over unresolved ionic (atomic) states is assumed.

The amplitude is given by (below we focus on the amplitude ${T}_s$
omitting its index)
\begin{equation}\label{amplitude_gen}
{T}=\langle{\bi k}_0\Phi_i^Z|{V}_i|\Psi^{-}_{f}({\bi k}_s,{\bi
k}_e)\rangle,
\end{equation}
where ${V}_{i}$ is the potential between the incident electron and
the atom. The initial asymptotic state $|{\bi
k}_0\Phi_i^Z\rangle\equiv|{\bi
k}_0\rangle\otimes|\Phi_i^Z\rangle$, where $|{\bi k}_0\rangle$ is
the plane wave state for the incident electron, satisfies the
Schr\"odinger equation
$$
({H}-{V}_i-E)|{\bi k}_0\Phi_i^Z\rangle=0,
$$
where ${H}$ is the full projectile-atom Hamiltonian and $E$ is
 the total energy:
$$E=E_s+E_e+\varepsilon_f=E_0+\varepsilon_i.$$ The total
scattering state $|\Psi^{-}_{f}({\bi k}_s,{\bi k}_e)\rangle$ takes
account of all interactions between the final-state fragments. It
satisfies the Schr\"odinger equation
$$
({H}-E)|\Psi^{-}_{f}({\bi k}_s,{\bi k}_e)\rangle=0
$$
and obeys the proper Coulomb asymptotics, whose specificity is due
to a long-range character of the interactions between the
final-state fragments.

\subsection{The plane wave Born series}
To avoid confusion, we will use a tilde for marking the
Lippmann-Schwinger analogs of the physical quantities that have
been introduced in the preceding subsection (such as the amplitude
$T$ and the total scattering state $|\Psi^{-}_{f}({\bi k}_s,{\bi
k}_e)\rangle$). In the context of the Lippmann-Schwinger formalism
which employs plane wave states for treating asymptotically free
particles, the total scattering state is sought as a solution to
the equation
\begin{eqnarray}
\label{ls} \fl |\tilde{\Psi}_f^-({\bi k}_s,{\bi
k}_e)\rangle&=&|{\bi k}_s{\bi
k}_e\Phi^{Z-1}_f\rangle+G_0^-(E)V|\tilde{\Psi}_f^-({\bi k}_s,{\bi
k}_e)\rangle, \qquad {V}={V}_{s}+{V}_{e}+{V}_{se},
\end{eqnarray}
where ${V}_{s}$, ${V}_{e}$, and ${V}_{se}$ are the electron-ion
and electron-electron potentials, respectively. The Green's
operator ${G}_0^-(E)$ is given by
\begin{equation}
\label{green_operator} 
{G}_0^-(E)=(E-{H}+{V}-\rmi0)^{-1}.
\end{equation}
The final asymptotic state $|{\bi k}_s{\bi
k}_e\Phi^{Z-1}_f\rangle\equiv|{\bi k}_s\rangle\otimes|{\bi
k}_e\rangle\otimes|\Phi^{Z-1}_f\rangle$, where $|{\bi k}_s\rangle$
and $|{\bi k}_e\rangle$ are, respectively, the plane wave states
for the scattered and ejected electrons, satisfies the
Schr\"odinger equation
$$
({H}-{V}-E)|{\bi k}_s{\bi k}_e\Phi^{Z-1}_f\rangle=0.
$$
\Eref{ls} can be presented in the equivalent form
\begin{equation}
\label{ls1} |\tilde{\Psi}_f^-({\bi k}_s,{\bi
k}_e)\rangle=[1+G^-(E)V]|{\bi k}_s{\bi k}_e\Phi^{Z-1}_f\rangle
\end{equation}
where $G^-(E)=(E-H-\rmi0)^{-1}$ is the full Green's operator.

%
%
Substitution of~(\ref{ls1}) into~(\ref{amplitude_gen}) generates
the plane wave perturbation series
\begin{eqnarray}
\label{born_series} \tilde{T}=\sum_{n=0}^\infty \tilde{T}^{(n)}
\qquad \mbox{where}~\tilde{T}^{(n)}=\langle{\bi
k}_0\Phi_i^Z|{V}_i[{G}_0^-(E){V}]^n|{\bi k}_s{\bi
k}_e\Phi^{Z-1}_f\rangle,
\end{eqnarray}
which is traditionally referred to as the Born series. Here the $n=0$
term amounts to the usual PWBA
\begin{equation}
\label{ls_fba} \tilde{T}^{(0)}\equiv T^{\rm PWBA}=\langle{\bi
k}_0\Phi_i^Z|{V}_i|{\bi k}_s{\bi k}_e\Phi^{Z-1}_f\rangle.
\end{equation}
All other (higher-order) Born terms $\tilde{T}^{(n)}$ and the sum
of these terms are divergent (see~\cite{popov81}
and~\ref{born_series_div} for details). These observations are not
novel and are due to a well-known fact that the plane wave states
do not obey the asymptotic conditions peculiar to break-up Coulomb
scattering. Thus, equation~\eref{born_series} gives an unphysical
result and hence it can not be employed for a perturbative
treatment of the correct, physical amplitude~\eref{amplitude_gen}.
In addition, the validity of the traditional PWBA
model~\eref{ls_fba} can not be based upon
equation~\eref{born_series}.

\subsection{The off-shell Born series}
The situation changes if we consider the Born
series~(\ref{born_series}) off the energy shell, i.e. when $E\neq
E_s+E_e+\varepsilon_f$. Setting
$E-E_s-E_e-\varepsilon_f=\Delta>0$, we have
\begin{equation}
\label{born_series_off_shell} \tilde{T}(\Delta)=\sum_{n=0}^\infty \tilde{T}^{(n)}(\Delta)\equiv T^{\rm
PWBA}+\sum_{n=1}^\infty \tilde{T}^{(n)}(\Delta) \qquad (\tilde{T}^{(n)}(0)\equiv \tilde{T}^{(n)}),
\end{equation}
where all the higher-order Born terms are finite~\cite{popov81} (see also~\ref{born_series_div}). If the value of the
off-shell parameter $\Delta$ approaches the on-shell case $\Delta=0$, the Born series~\eref{born_series_off_shell}
exhibits a typical Coulomb singularity
$\tilde{T}(\Delta)\propto\Delta^{-\rmi\eta}$~\cite{latypov93,shablov02,faddeev2}, where $\eta$ is a total Sommerfeld
parameter:
$$
\eta=\eta_s+\eta_e+\eta_{se} \qquad
(\eta_{s}=-k_{s}^{-1},~\eta_{e}=-k_{e}^{-1},~\mbox{and}~\eta_{se}=|{\bi
k}_s-{\bi k}_e|^{-1}).
$$

The physical amplitude~\eref{amplitude_gen} is derived
from~\eref{born_series_off_shell} in the following
manner~\cite{shablov99,shablov02,faddeev2}:
\begin{equation}
\label{amplitude_exact} {T}=\frac{\exp(-\case12\pi\eta-\rmi
A)}{\Gamma(1+\rmi\eta)}
\lim_{\Delta\rightarrow0}\Delta^{\rmi\eta}\tilde{T}(\Delta),
\end{equation}
where
$$
A=\eta_s\ln(2k_s^2)+\eta_e\ln(2k_e^2) +\eta_{se}\ln|{\bi k}_s-{\bi
k}_e|^2
$$
is the Dollard phase~\cite{dollard64}. Note that in the on-shell limit $\Delta\to0$ the divergent factor
$\Delta^{\rmi\eta}$ compensates for the singularity of $\tilde{T}(\Delta)$. In the next section we show how the
calculation scheme based on equation~\eref{amplitude_exact} can be implemented in practice through regularization of
the Born series.

\section{Born approximations}
\label{regproc}
%

%
%
For taking the on-shell limit $\Delta\to0$ in
equation~(\ref{amplitude_exact}), it is convenient to have
$\Delta^{-\rmi\eta}$ factored out in the off-shell Born
series~\eref{born_series_off_shell}. This can be fulfilled by
means of a regularization procedure (see~\ref{reg_proc}). As a
result, we obtain the amplitude~\eref{amplitude_exact} in the
factorized form
\begin{equation}
\label{amplitude_exact_1} {T}=\mathcal{R}\tilde{T}_\mathcal{R},
\qquad
\tilde{T}_\mathcal{R}=\sum_{n=0}^\infty\tilde{T}_\mathcal{R}^{(n)}\equiv
T^{\rm PWBA}+\sum_{n=1}^\infty\tilde{T}_\mathcal{R}^{(n)},
\end{equation}
where $\tilde{T}_\mathcal{R}^{(n)}$ is the regularized on-shell
Born term and $\mathcal{R}$ is a regularization
function\footnote[1]{Hereafter $\mathcal{R}$ refers to an
arbitrary regularization procedure/function unless otherwise
specified.}, such that $\mathcal{R}=1$ if
$\eta_s=\eta_e=\eta_{se}=0$.

In contrast to \eref{born_series}, the
result~(\ref{amplitude_exact_1}) allows to develop a perturbative
treatment of the physical amplitude. However, when truncating the
regularized on-shell Born series $\tilde{T}_\mathcal{R}$ to a
finite number of terms, there is an uncertainty associated with
the choice of a regularization procedure. Namely, while
$$\mathcal{R}\tilde{{T}}_\mathcal{R}=\mathcal{R}'\tilde{{T}}_{\mathcal{R}'},$$
where $\mathcal{R}'$ and $\tilde{{T}}_{\mathcal{R}'}$ are due to
an alternative regularization procedure (see~\ref{reg_proc} for
details), in general
$$\mathcal{R}'\sum_{n=0}^N\tilde{{T}}_{\mathcal{R}'}^{(n)}\neq\mathcal{R}\sum_{n=0}^N\tilde{{T}}_\mathcal{R}^{(n)}.$$
In particular, the lowest-order approximation ($N=0$) to the
amplitude~(\ref{amplitude_exact_1}) assumes the form
\begin{equation}
\label{FBA_present} T^{\rm
PWBA}_\mathcal{R}=\mathcal{R}\tilde{{T}}_\mathcal{R}^{(0)}\equiv\mathcal{R}T^{\rm
PWBA}
\end{equation}
that depends on the choice of a regularization procedure/function.
These observations are reminiscent of the situation that one
encounters in quantum electrodynamics, when regularizing a series
of Feynman's diagrams. Drawing an analogy with that situation, one
can speak of a renormalization group which is formed by the
regularization procedures in the present case. Therefore we will
refer to equation~(\ref{FBA_present}) as a renormalized PWBA
(RPWBA).

In this work we develop the Born series for the physical
amplitude~\eref{amplitude_gen} in a manner similar to
equation~(\ref{born_series}), that is
\begin{eqnarray}
\label{amplitude_exact_correct} T=\sum_{n=0}^\infty T^{(n)},
\end{eqnarray}
where $T^{(n)}$ is given by the $n$th-order term of the Maclaurin
series expansion of the exact amplitude $T$ with respect to the
two-particle Sommerfeld parameters $\eta_s$, $\eta_e$, and
$\eta_{se}$. It can be deduced from~\eref{amplitude_exact_1} that
\begin{equation}\label{born_term_correct}
T^{(n)}=\tilde{T}^{(n)}_{\mathcal{R}=1},
\end{equation}
where $\tilde{T}^{(n)}_{\mathcal{R}=1}$ corresponds to the
specific regularization procedure which yields $\mathcal{R}=1$ for
any values of $\eta_s$, $\eta_e$, and $\eta_{se}$
(see~\ref{reg_proc}).



According to~\eref{born_term_correct}, the lowest-order term
of~\eref{amplitude_exact_correct} amounts to the usual PWBA
result~\eref{ls_fba}:
\begin{equation}\label{pwba}
T^{(0)}\equiv T^{\rm PWBA}=\langle{\bi k}_0\Phi_i^Z|{V}_i|{\bi
k}_s{\bi k}_e\Phi^{Z-1}_f\rangle.
\end{equation}
The plane wave second Born approximation (PWB2), which is usually
of practical value for estimating the applicability of the PWBA,
is then given by
\begin{eqnarray}
\label{SBA_correct}T^{\rm PWB2}=T^{(0)}+T^{(1)}=T^{\rm
PWBA}+\tilde{T}^{(1)}_{\mathcal{R}=1}.
\end{eqnarray}
Using~\eref{amplitude_exact}, this is equivalent to
\begin{equation}
\label{SBA_correct_1}\fl T^{\rm
PWB2}=\left(1-\frac{\pi\eta}{2}-\rmi A+\rmi\gamma\eta\right)T^{\rm
PWBA}+\lim_{\Delta\rightarrow0}[\tilde{T}^{(1)}(\Delta)+\rmi\eta
T^{\rm PWBA}\ln\Delta],
\end{equation}
where $\gamma=0.577216$ is the Euler constant.

\section{Impulse approximations}
\label{impapprox}
\subsection{The usual theory}
Near the Bethe ridge the (e,2e) process can be modelled as a
binary encounter between the projectile electron and the electron
that is ejected from the atom~\cite{weigold_book}. The PWIA theory
formulates this picture mathematically. Within the
Lippmann-Schwinger approach~\eref{born_series} it is expressed as
follows:
\begin{equation}
\label{ls_pwia1}\fl \tilde{T}^{\rm PWIA}=\sum_{n=0}^\infty
\tilde{T}^{(n)}_{se}, \qquad
\mbox{where}~\tilde{T}^{(n)}_{se}=\langle{\bi
k}_0\Phi_i^Z|{V}_{se}[{G}_0^-(E){V}_{se}]^n|{\bi k}_s{\bi
k}_e\Phi^{Z-1}_f\rangle.
\end{equation}
As in the case of~\eref{born_series}, the $n\geq1$ terms and the
sum of these terms are divergent. We can remedy this defect in
order to obtain a correct, convergent counterpart
of~\eref{ls_pwia1}. For this purpose we use the
result~\eref{amplitude_exact_1}, assuming that $V_i=V=V_{se}$ and,
accordingly, $\eta_s=\eta_e=0$. Thus one derives the physical
counterpart of~\eref{ls_pwia1} as
\begin{equation}
\label{ems_pwia}{T}^{\rm PWIA}=\mathcal{R}_{se}\tilde{t}_\mathcal{R}F_{if}({\bi q})=\tau
F_{if}({\bi q})\qquad (\mathcal{R}_{se}=\mathcal{R}|_{\eta_s=\eta_e=0}),
\end{equation}
where ${\bi q}={\bi k}_s+{\bi k}_e-{\bi k}_0$ is opposite to the
recoil ion momentum and $F_{if}({\bi
q})=\langle\Phi_i^Z|\Phi_f^{Z-1}{\bi q}\rangle$ is the so-called
structure amplitude~\cite{weigold_book}. $\tilde{t}_\mathcal{R}$
is the regularized Lippmann-Schwinger half-on-shell amplitude for
ee-scattering~\cite{popov81} and $\tau$ is the exact half-on-shell
ee-scattering amplitude~\cite{ford64,hostler64}:
$$
\tau=\tau\{{\bi k}_0-\case{1}{2}({\bi k}_s+{\bi
k}_e),\case{1}{2}({\bi k}_s-{\bi k}_e);[\case{1}{2}({\bi k}_s-{\bi
k}_e)]^2\}.
$$
The result~\eref{ems_pwia} is irrespective of the choice of a
regularization procedure, since
$\mathcal{R}_{se}\tilde{t}_\mathcal{R}=\mathcal{R}_{se}'\tilde{t}_{\mathcal{R}'}$.
Note that equation~\eref{ems_pwia} is a corner-stone of the usual
PWIA theory of (e,2e) reactions on atoms~\cite{weigold_book,neud}.

\subsection{A renormalized theory}
While there is nothing wrong in the above derivation of the
traditional PWIA result, it can be noticed that at the starting
point~\eref{ls_pwia1} we have focused on the purely
electron-electron part of~\eref{born_series} and thereby we have
ignored the remaining part which is also divergent.
Methodologically, one should apply the binary-encounter
approximation to the regularized Lippmann-Schwinger amplitude
$\tilde{T}_\mathcal{R}$ (see~\eref{amplitude_exact_1}) which, in
contrast to~\eref{born_series}, is free of divergences. The
electron-electron component of $\tilde{T}_\mathcal{R}$ is given by
the regularized Lippmann-Schwinger amplitude
$\tilde{t}_\mathcal{R}$ for ee-scattering (see~\eref{ems_pwia}).
Using~\eref{amplitude_exact_1} and~\eref{ems_pwia}, the PWIA to
the physical amplitude $T$ is then given by
\begin{equation}
\label{amplitude_pwia} T^{\rm
PWIA}_\mathcal{R}=\mathcal{R}\tilde{t}_\mathcal{R}F_{if}({\bi
q})=(\mathcal{R}/\mathcal{R}_{se})\tau F_{if}({\bi q}).
\end{equation}
Since in general
$\mathcal{R}/\mathcal{R}_{se}\neq\mathcal{R}'/\mathcal{R}_{se}'$,
the PWIA amplitude~\eref{amplitude_pwia} is determined by the
choice of a regularization procedure. In particular, the specific
case $\mathcal{R}/\mathcal{R}_{se}=1$ amounts to the traditional
PWIA~(\ref{ems_pwia}). As far as the choice of a regularization
procedure is a matter of taste, equation~\eref{amplitude_pwia}
offers an infinite number of alternatives. This fact has the
following general consequence: there is no \emph{a priori} PWIA
model in the case of (e,2e) reactions on atoms. By analogy with
the RPWBA case~\eref{FBA_present}, the
approximation~\eref{amplitude_pwia} will be referred to as a
renormalized PWIA (RPWIA).

\subsection{TDCS and the Gamow factor}
\label{gamow_factor}
Within the traditional PWIA model~\eref{ems_pwia} the
TDCS~\eref{tdcs} is expressed as~\cite{weigold_book}
\begin{eqnarray}
\label{tdcs_ems} \fl \frac{\rmd^3\sigma^{\rm
PWIA}}{\rmd\Omega_s\rmd\Omega_e\rmd E_e}=\frac{k_sk_e}{2\pi^3k_0}
\frac{\mathcal{G}(\eta_{se})}{|{\bi k}_0-{\bi k}_s|^4}\nonumber\\
\times\left[1+\frac{|{\bi k}_0-{\bi k}_s|^4}{|{\bi k}_0-{\bi
k}_e|^4}-\frac{|{\bi k}_0-{\bi k}_s|^2}{|{\bi k}_0-{\bi
k}_e|^2}\cos\left(\eta_{se}\ln\frac{|{\bi k}_0-{\bi k}_s|}{|{\bi
k}_0-{\bi k}_e|}\right)\right]\nonumber\\
\times \sum\nolimits^{({\rm av})}|F_{if}({\bi q})|^2,
\end{eqnarray}
where $\sum\nolimits^{({\rm av})}$ denotes the average over
initial-state and sum over final-state degeneracies,
%
and $\mathcal{G}(\eta_{se})$ is the so-called Gamow
factor~\cite{gamow28}:
\begin{equation}
\label{gamow_traditional}
\mathcal{G}(\eta_{se})=|\exp(-\case12\pi\eta_{se})\Gamma(1-\rmi\eta_{se})|^2=\frac{2\pi\eta_{se}}{\rme^{2\pi\eta_{se}}-1}.
\end{equation}
In the case of the RPWIA model~\eref{amplitude_pwia} the
expression for TDCS depends on the explicit form of the
regularization function $\mathcal{R}$. In this work we inspect the
following form:
\begin{equation}\label{reg_function_present}
\fl \mathcal{R}=\exp(-\case12\pi\eta)\Gamma(1-\rmi\eta)\qquad \mbox{and
consequently}~\mathcal{R}_{se}=\exp(-\case12\pi\eta_{se})\Gamma(1-\rmi\eta_{se}).
\end{equation}
Substitution of~\eref{amplitude_pwia}
and~\eref{reg_function_present} into~\eref{tdcs} yields
\begin{eqnarray}
\label{tdcs_pwia_present} \fl \frac{\rmd^3\sigma^{\rm
RPWIA}}{\rmd\Omega_s\rmd\Omega_e\rmd E_e}=\frac{k_sk_e}{2\pi^3k_0}
\frac{\mathcal{G}(\eta)}{|{\bi k}_0-{\bi k}_s|^4}\nonumber\\
\times\left[1+\frac{|{\bi k}_0-{\bi k}_s|^4}{|{\bi k}_0-{\bi
k}_e|^4}-\frac{|{\bi k}_0-{\bi k}_s|^2}{|{\bi k}_0-{\bi
k}_e|^2}\cos\left(\eta_{se}\ln\frac{|{\bi k}_0-{\bi k}_s|}{|{\bi
k}_0-{\bi k}_e|}\right)\right]\nonumber\\
\times \sum\nolimits^{({\rm av})}|F_{if}({\bi q})|^2,
\end{eqnarray}
with the modified Gamow factor
\begin{equation}
\label{gamow_new}
\mathcal{G}(\eta)=\frac{2\pi\eta}{\rme^{2\pi\eta}-1}.
\end{equation}
%

As can be deduced, the RPWIA result~\eref{tdcs_pwia_present}
differs from the traditional PWIA one~\eref{tdcs_ems} only in the
expression for the Gamow factor. In contrast
to~\eref{gamow_traditional}, the modified Gamow
factor~\eref{gamow_new} treats the final-state particle pairs on
equal footing. It should be noted that the same result
as~\eref{gamow_new} is obtained using the first-order model
introduced by Shablov~\etal~\cite{shablov99} on the basis of the
formalism of regularization Coulomb operators, and it can be
traced down to the idea of effective charges proposed by
Peterkop~\cite{peterkop_book}. The influences of the traditional
and modified Gamow factors on TDCS are examined in the next
section, where we present the corresponding numerical results.

Using the RPWBA model~\eref{FBA_present} and
equation~\eref{reg_function_present}, in the binary-encounter
approximation ($V_i=V_{se}$) we get
\begin{eqnarray}
\label{tdcs_rba}
\frac{\rmd^3\sigma^{\rm RPWBA}}{\rmd\Omega_s\rmd\Omega_e\rmd
E_e}=\mathcal{G}(\eta)\frac{\rmd^3\sigma^{\rm
PWBA}}{\rmd\Omega_s\rmd\Omega_e\rmd E_e},
\end{eqnarray}
where the traditional PWBA result is given by~\cite{weigold_book}
\begin{eqnarray}
\label{tdcs_pwba}\fl \frac{\rmd^3\sigma^{\rm
PWBA}}{\rmd\Omega_s\rmd\Omega_e\rmd
E_e}=\frac{k_sk_e}{2\pi^3k_0}\frac{1}{|{\bi k}_0-{\bi
k}_s|^4}\left[1+\frac{|{\bi k}_0-{\bi k}_s|^4}{|{\bi k}_0-{\bi
k}_e|^4}-\frac{|{\bi k}_0-{\bi k}_s|^2}{|{\bi k}_0-{\bi
k}_e|^2}\right]\sum\nolimits^{({\rm av})}|F_{if}({\bi q})|^2.
\end{eqnarray}
It can be noticed that in symmetric kinematics ($E_s=E_e$,~$|{\bi k}_0-{\bi k}_s|=|{\bi k}_0-{\bi k}_e|$), which is
usually the case of the EMS experiments (see~\cite{weigold_book,watanabe05,ren05} and references therein), the TDCS
given by equation~\eref{tdcs_rba} is identical to that given by equation~\eref{tdcs_pwia_present}.

\section{Numerical realization}
\label{numres}
In this section we present numerical results for an archetypical
case, namely the (e,2e) reaction on a hydrogen atom. The symmetric
setup $E_s=E_e$ and $\theta_s=\theta_e=45^\circ$ is inspected,
where the polar electron angles $\theta_s$ and $\theta_e$ are
measured with respect to the direction of the incident electron
momentum. In the considered setup the TDCS is usually studied as a
function of $q=|{\bi q}|$ (see~\eref{ls_pwia1}) whose value is
varied in noncoplanar geometry by varying the value of the
relative azimuthal angle $\Delta\phi_{se}=\phi_s-\phi_e$, where
$\phi_s$ ($\phi_e$) is the azimuthal angle of the scattered
(ejected) electron. The minimal value of $q$ corresponds then to
the symmetric coplanar case ($|\Delta\phi_{se}|=\pi$), where the
incident and outgoing electron momenta are in the same plane.

We focus on high energy values ($E_s=E_e\gtrsim1$~keV) which
closely meet the Bethe-ridge and binary-encounter criteria. In the
absence of the corresponding EMS measurements, particularly those
performed on an absolute scale, the results of the traditional and
renormalized first-order treatments are compared with those of the
PWB2 calculations and those of the Brauner-Briggs-Klar (BBK)
model~\cite{bbk}, a representative of nonperturbative treatments.
The PWB2 calculations have been carried out in accordance
with~\eref{SBA_correct_1}, where the on-shell limit has been taken
analytically using the regularization procedure described
in~\ref{reg_proc}. The regularized PWB2 integrals have been
performed numerically following the method developed
in~\cite{pasha05}.

In the BBK model the final state in~\eref{amplitude_gen} is given
by
\begin{equation}
\label{bbk}\fl|\Psi^{-}_{\rm BBK}({\bi k}_s,{\bi
k}_e)\rangle=|\varphi^{-}({\bi
k}_s)\rangle\otimes|\varphi^{-}({\bi
k}_e)\rangle\otimes|\chi^{-}({\bi k}_{se})\rangle \qquad ({\bi
k}_{se}=\case12({\bi k}_s-{\bi k}_e)),
\end{equation}
where $|\varphi^{-}({\bi k}_{s})\rangle$ and $|\varphi^{-}({\bi
k}_{e})\rangle$ are the Coulomb waves describing outgoing
electrons moving in the field of the proton. The electron-electron
correlation factor $|\chi^{-}({\bi k}_{se})\rangle$ is determined
through
$$
|{\bi k}_{s}\rangle\otimes|{\bi k}_{e}\rangle\otimes|\chi^{-}({\bi
k}_{se})\rangle=|\psi^{-}({\bi k}_{s},{\bi k}_{e})\rangle,
$$
where $|\psi^{-}({\bi k}_{s},{\bi k}_{e})\rangle$ is the exact
scattering state in the absence of the electron-proton
interactions. The wave function~\eref{bbk} has the correct
asymptotic behavior and gives correct results in the limiting
situations, where (i) the charge of one of the final-state
particles is switched off and (ii) the electron-electron
interaction is absent. Note that the BBK model yields the
regularization function~\eref{reg_function} as
\begin{equation}
\label{reg_function_BBK}R_{\rm BBK}=\exp(-\case12\pi\eta)\Gamma(1-\rmi\eta_s)
\Gamma(1-\rmi\eta_e)\Gamma(1-\rmi\eta_{se}).
\end{equation}
The corresponding Gamow factor is then given by
%
\begin{equation}
\label{gamow_BBK} \mathcal{G}_{\rm
BBK}=\mathcal{G}(\eta_{s})\mathcal{G}(\eta_{e})\mathcal{G}(\eta_{se})=
\frac{2\pi\eta_{s}}{\rme^{2\pi\eta_{s}}-1}\,\frac{2\pi\eta_{e}}{\rme^{2\pi\eta_{e}}-1}
\,\frac{2\pi\eta_{se}}{\rme^{2\pi\eta_{se}}-1}.
\end{equation}
Like~\eref{gamow_new}, this Gamow factor also treats the
final-state particle pairs on equal footing.

The numerical calculations using the traditional PWBA and PWIA
models have been performed in accordance with
equations~\eref{tdcs_pwba} and~\eref{tdcs_ems}, respectively. As
remarked in subsection~\ref{gamow_factor}, in the case of
symmetric kinematics the RPWBA~\eref{tdcs_rba} and
RPWIA~\eref{tdcs_pwia_present} models are equivalent and therefore
below the corresponding numerical results referred to as
RPWBA/RPWIA. For a hydrogen target, the structure factor entering
equations~\eref{tdcs_ems},~\eref{tdcs_pwia_present},~\eref{tdcs_rba},
and~\eref{tdcs_pwba} is
\begin{equation}
\label{structure_factor}\sum\nolimits^{({\rm av})}|F_{if}({\bi
q})|^2=|\varphi_{1s}(q)|^2,
\end{equation}
where $\varphi_{1s}(q)$ is the $1s$ state momentum-space wave
function.

\subsection{Numerical results and discussion}

\begin{figure}
\begin{center}
\epsfxsize=12cm \epsfbox{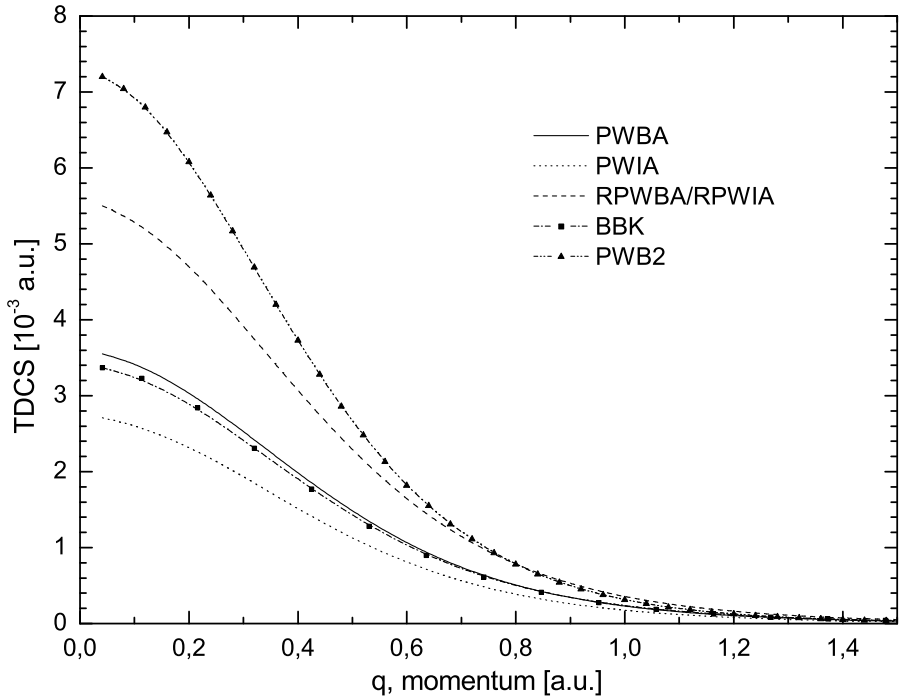}
\end{center}
\caption{\label{fig1}TDCS as a function of the absolute value of
the recoil ion momentum in the symmetric noncoplanar kinematics
($E_s=E_e=1000$~eV, $\theta_s=\theta_e=45^\circ$) of the recent
(e,2e) experiments~\cite{watanabe05}.}
\end{figure}

Figure~\ref{fig1} shows the numerical results for the symmetric
noncoplanar kinematics utilized in the recent (e,2e) measurements
on helium~\cite{watanabe05}. It can be seen that the RPWBA/RPWIA
results are substantially larger in magnitude than the PWIA ones,
which are the smallest in magnitude. This feature indicates an
appreciable role of the choice of the Gamow factor in the
kinematics under consideration. Interestingly, the PWBA and BBK
results are close to each other both in magnitude and in shape.
Though the BBK model is not exact, it takes into account those
higher-order effects that are entirely neglected by the PWBA
treatment. One might thus conclude that the higher-order
contributions to the TDCS are subsidiary in the present case.
However, this conclusion needs experimental verification, since
marked discrepancies between the PWBA and PWB2 results are
observed in figure~\ref{fig1}. It should be remarked that the
developed renormalized first-order theories give an opportunity to
fit the exact TDCS by the proper choice of the Gamow factor. For
example, setting (cf equation~\eref{gamow_new})
$$
G(\eta)=1
$$
in equations~\eref{tdcs_pwia_present} and~\eref{tdcs_rba}, one
obtains the traditional PWBA result and thus, as can be seen in
figure~\ref{fig1}, rather well reproduces the BBK results for the
present kinematics. And the PWB2 results are satisfactorily
reproduced in magnitude using the modified Gamow
factor~\eref{gamow_new}.

\begin{figure}
\begin{center}
\epsfxsize=12cm \epsfbox{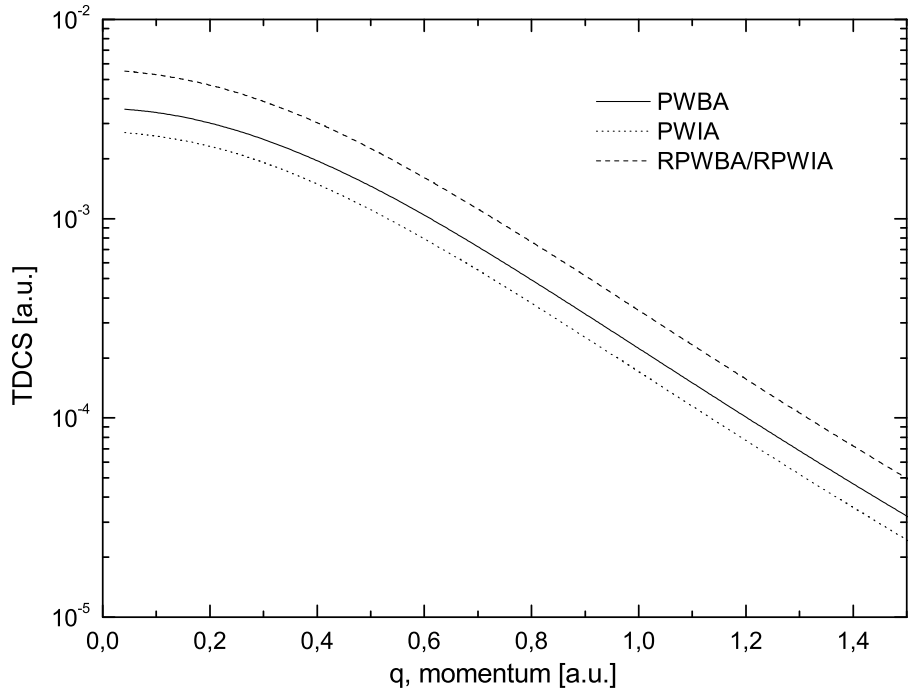}
\end{center}
\caption{\label{fig2}The first-order results for the TDCS. The
kinematics is the same as in figure~\ref{fig1}.}
\end{figure}
\begin{figure}
\begin{center}
\epsfxsize=12cm \epsfbox{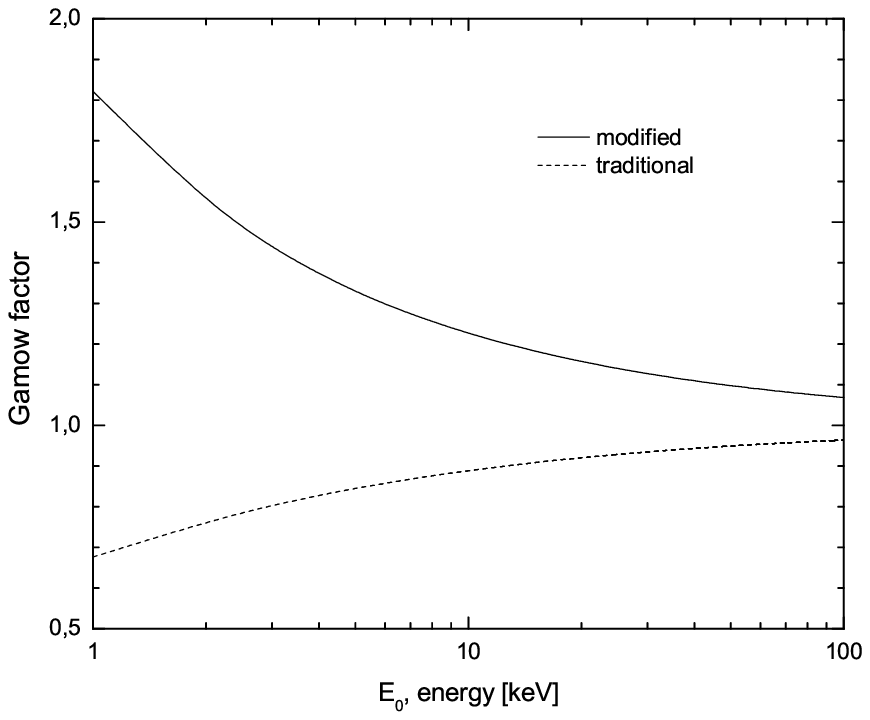}
\end{center}
\caption{\label{fig3}The traditional~\eref{gamow_traditional} and
modified~\eref{gamow_new} Gamow factors as functions of the
incident energy $E_0$ in the symmetric coplanar case ($E_s=E_e$,
$\theta_s=\theta_e=45^\circ$, and $|\Delta\phi_{se}|=\pi$).}
\end{figure}

In figure~\ref{fig2} the traditional PWBA and PWIA and the
RPWBA/RPWIA values are presented on a logarithmic scale. It can be
seen that all three models exhibit practically identical shapes
for the TDCS. This feature is due to the almost constancy of the
Sommerfeld parameters $\eta_{se}$ and $\eta$, and hence of the
Gamow factors~(\ref{gamow_traditional}) and~(\ref{gamow_new}), in
the involved kinematical region. It should be noted that the value
of the Gamow factor~\eref{gamow_BBK} is nearly the same as that
of~\eref{gamow_new}. This observation can be explained by the
feature that the first-order expansions of~\eref{gamow_new}
and~\eref{gamow_BBK} with respect to $\eta_s$, $\eta_e$, and
$\eta_{se}$ are identical. Another important observation is that,
in contrast to the traditional Gamow
factor~(\ref{gamow_traditional}), the modified Gamow
factor~(\ref{gamow_new}) enhances the magnitude of the TDCS with
respect to that in the conventional PWBA case. At the same time,
all three first-order models are practically equivalent for
description of the corresponding symmetric noncoplanar
measurements provided that the latter are performed on a relative
scale (see, for instance,~\cite{lohmann81}). This conclusion
follows from the fact that in the case of the inspected
first-order models the shape of the TDCS (the so-called momentum
profile~\cite{weigold_book,watanabe05,ren05}) is almost fully
determined by the structure factor~\eref{structure_factor}.

The effect of the Gamow factor on the magnitude of TDCS in the
considered geometry is shown in figure~\ref{fig3}. As can be
deduced from the figure, the RPWBA/RPWIA and the traditional PWIA
values for the TDCS rather slowly converge to each other and
ultimately to the traditional PWBA value upon the increase of the
incident electron energy $E_0$. For $E_0\sim10$~keV, the
RPWBA/RPWIA results are larger in magnitude than the traditional
PWIA ones by a factor of $\sim1.5$, and even in the region
$E_0\sim100$~keV, where one might expect relativistic effects to
come on the scene, the relative difference between the results in
magnitude is about 10\%. These findings are in discord with an
intuitive physical picture which assumes practical equivalence of
the first-order models in the kinematics that approaches the
classical ridge of a billiard-ball collision~\cite{neud}. The
discrepancy can be explained by the following factors: (i) the
long-range Coulomb forces between the colliding electrons, as
opposed to the contact-like forces between classical billiard
balls, and (ii) the presence of the Coulomb field of the ion (this
factor is relevant only to the RPWBA/RPWIA model).

\section{Summary and conclusions}
\label{concl}
In summary, we have considered the electron-atom ionization
process at large energy-momentum transfer and near the Bethe
ridge. Proceeding from the plane wave Lippmann-Schwinger
amplitude, which diverges on the energy shell, we have regularized
the corresponding Born series. On this basis we have developed the
correct, physical Born treatment whose lowest-order term amounts
to the conventional PWBA. The RPWBA model has been formulated,
which depends on the employed regularization procedure. We have
shown that the PWIA can not be uniquely determined and therefore
we have introduced the RPWIA model, in which the Gamow factor is
determined by the choice of a regularization procedure. The
numerical results for the symmetric noncoplanar kinematics have
been presented. It has been demonstrated that even at high
electron energies ($\sim10$~keV) the magnitude of the TDCS is very
sensitive to the choice of the Gamow factor.

We might expect the results of this work to be primarily important
for theoretical treatments of absolute (e,2e) measurements in the
nearly Bethe-ridge kinematics at large energy-momentum transfer
and for further development of the EMS method~\cite{weigold_book}.
The present theoretical consideration can be generalized to the
case of ionization of an atom by a charged-particle impact, for
example, to the cases of positron- and proton-atom ionization.
Using the formulated RPWBA and RPWIA models with a properly
modified Gamow factor, one can efficiently take into account the
higher-order effects ignored by the traditional PWBA and PWIA
models. In this connection, it should be noted that a consistent,
rigorous treatment of the higher-order contributions to the
ionization amplitude is realized by means of the developed plane
wave Born series~\eref{amplitude_exact_correct}, which has an
apparent advantageous feature: the value of any of its terms is,
by definition, irrespective of the choice of a regularization
procedure.

\ack
We are grateful to Vladimir L. Shablov and Ochbadrakh
Chuluunbaatar for useful discussions. We would like to thank
Amulya Roy for helpful comments and remarks.

%
\appendix
\section{Divergencies of the Born series}\label{born_series_div}
To elucidate the origin of divergencies, let us examine the
electron-electron part of the $n=1$ term in~\eref{born_series}
\begin{eqnarray}\label{example}\fl \tilde{T}^{(1)}_{se}=
\int\frac{\rmd{\bi p}_s}{(2\pi)^3} \frac{\rmd{\bi p}_e}{(2\pi)^3}
\frac{\langle{\bi k}_0\Phi_i^Z|{V}_i|{\bi p}_s{\bi
p}_e\Phi^{Z-1}_f\rangle}{E-p_s^2/2-p_e^2/2-\varepsilon_f-\rmi0}
\langle{\bi p}_s{\bi p}_e|{V}_{se}|{\bi k}_s{\bi
k}_e\rangle\nonumber\\\bs \lo =\int\frac{\rmd{\bi p}}{(2\pi)^3}
\frac{\langle{\bi k}_0\Phi_i^Z|{V}_i|{\bi k}_s-{\bi p},{\bi
k}_e+{\bi p},\Phi^{Z-1}_f\rangle}{({\bi k}_s-{\bi k}_e)\cdot{\bi
p}-p^2-\rmi0}\frac{4\pi}{p^2}.
\end{eqnarray}
It can be seen that the integrand has a pronounced singularity at
the point ${\bi p}=0$ which physically corresponds to elastic
rescattering in the forward direction. This feature makes the
integral~(\ref{example}) divergent. The same argument applies to
the electron-ion parts of the $n=1$ term, since for the static
matrix elements of $V_s$ and $V_e$ we have:
$$
\langle{\bi k}_s\pm{\bi p},\Phi^{Z-1}_f|{V}_{s}|{\bi
k}_s\Phi^{Z-1}_f\rangle=\langle{\bi k}_e\pm{\bi
p},\Phi^{Z-1}_f|{V}_{e}|{\bi
k}_e\Phi^{Z-1}_f\rangle\simeq-\frac{4\pi}{p^2} \qquad (p\to0).
$$
Clearly, in the case $n\geq2$ we encounter even
stronger divergencies because the corresponding Born terms contain
multiple elastic rescattering in the forward direction.

In the case of~\eref{born_series_off_shell}, the electron-electron
part of the $n=1$ term takes the form (cf~(\ref{example}))
\begin{eqnarray}\label{example_1} \tilde{T}^{(1)}_{se}(\Delta)= \int\frac{\rmd{\bi p}}{(2\pi)^3}\frac{\langle{\bi
k}_0\Phi_i^Z|{V}_i|{\bi k}_s-{\bi p},{\bi k}_e+{\bi
p},\Phi^{Z-1}_f\rangle }{\Delta+({\bi k}_s-{\bi k}_e)\cdot{\bi
p}-p^2-\rmi0}\frac{4\pi}{p^2},
\end{eqnarray}
where the singularity in the integrand at the point ${\bi p}=0$ is removed (note that $\rmd{\bi
p}=p^2\rmd p\,\rmd\Omega_p$) and thereby the integral does not diverge. Specifically, in the case
$\Delta\to0$ we have~\cite{popov81}
$$
\tilde{T}^{(1)}_{se}(\Delta)\sim-\rmi\eta_{se}T^{\rm
PWBA}\ln\Delta.
$$
\section{Regularization procedure}\label{reg_proc}
Below we describe a possible recipe for factoring out
$\Delta^{-\rmi\eta}$ in equation~\eref{born_series_off_shell} and
then taking the on-shell limit in equation~\eref{amplitude_exact}.
It consists in presenting the Green's
operator~(\ref{green_operator}) in the form
\begin{eqnarray}
\label{green_operator1} \fl {G}_0^-(E)=\sum_{f'}\int\frac{\rmd
{\bi p}_s}{(2\pi)^3}\frac{\rmd {\bi p}_e}{(2\pi)^3}\frac{|{\bi
p}_s{\bi p}_e\Phi^{Z-1}_{f'}\rangle\langle{\bi p}_s{\bi
p}_e\Phi^{Z-1}_{f'}|}{E-p_s^2/2-p_e^2/2-\varepsilon_{f'}-\rmi0}
={\mathcal{G}}_0^-(E)+{\mathcal{F}}_0^-(E),
\end{eqnarray}
where ${\mathcal{G}}_0^-(E)$ is the regularized Green's operator
and
\begin{equation}
\label{div_part} {\mathcal{F}}_0^-(E)= \sum_{f'}\int\frac{\rmd
{\bi p}_s}{(2\pi)^3}\frac{\rmd {\bi p}_e}{(2\pi)^3}\frac{|{\bi
k}_s{\bi k}_e\Phi^{Z-1}_{f}\rangle\langle{\bi p}_s{\bi
p}_e\Phi^{Z-1}_{f'}|}{E-p_s^2/2-p_e^2/2-\varepsilon_{f'}-\rmi0}
\end{equation}
is the Green's operator component which is responsible for
divergencies. Using~\eref{green_operator1} and~\eref{div_part}, we
obtain the off-shell Born series \eref{born_series_off_shell} in
the factorized form
\begin{eqnarray}\label{reg1}
\tilde{T}(\Delta)={P}_R(\Delta)\tilde{T}_R(\Delta), 
\end{eqnarray}
where
$$
\tilde{T}_R(\Delta)=\sum_{n=0}^\infty\tilde{T}^{(n)}_R(\Delta)\equiv
T^{\rm PWBA}+\sum_{n=1}^\infty\tilde{T}^{(n)}_R(\Delta),
$$
$$
{P}_R(\Delta)=\sum_{n=0}^\infty{P}^{(n)}_R(\Delta)\equiv1+\sum_{n=1}^\infty{P}^{(n)}_R(\Delta),
$$
with
%
$$\tilde{T}^{(n)}_R(\Delta)=\langle{\bi
k}_0\Phi_i^Z|{V}_i[{\mathcal{G}}_0^-(E){V}]^n|{\bi k}_s{\bi
k}_e\Phi^{Z-1}_f\rangle,
$$
%
%
$$
{P}^{(n)}_R(\Delta)=\sum_{f'}\int\frac{\rmd {\bi
p}_s}{(2\pi)^3}\frac{\rmd {\bi p}_e}{(2\pi)^3}\langle{\bi p}_s{\bi
p}_e\Phi_{f'}^{Z-1}|[{G}_0^-(E){V}]^n|{\bi k}_s{\bi
k}_e\Phi^{Z-1}_f\rangle.
$$
Taking into account that
$$ 1+\sum_{n=1}^\infty[{G}_0^-(E){V}]^n=1+{G}^-(E){V}
$$
and using~\eref{ls1}, we get
\begin{equation}\label{reg3}
\lim_{\Delta\to0}{P}_R(\Delta)=\sum_{f'}\int\frac{\rmd {\bi
p}_s}{(2\pi)^3}\frac{\rmd {\bi p}_e}{(2\pi)^3}\langle{\bi p}_s{\bi
p}_e\Phi_{f'}^{Z-1}|\tilde\Psi^{-}_{f}({\bi k}_s,{\bi
k}_e)\rangle.
\end{equation}
The on-shell limit~\eref{reg3} does not exist, since the
Lippmann-Schwinger total scattering state
$|\tilde\Psi^{-}_{f}({\bi k}_s,{\bi k}_e)\rangle$ is not physical.
Using the results of Shablov \etal~\cite{shablov99,shablov02}, we
deduce that
\begin{eqnarray}\label{reg4}
\fl {P}_R(\Delta\to0)=\Delta^{-\rmi\eta}\exp(\case12\pi\eta+\rmi
A)\Gamma(1+\rmi \eta)\nonumber\\\bs
 \times\sum_{f'}\int\frac{\rmd {\bi p}_s}{(2\pi)^3}\frac{\rmd
{\bi p}_e}{(2\pi)^3}\langle{\bi p}_s{\bi
p}_e\Phi_{f'}^{Z-1}|\Psi^{-}_{f}({\bi k}_s,{\bi k}_e)\rangle.
\end{eqnarray}
Inserting~\eref{reg1} into~\eref{amplitude_exact} and
using~\eref{reg4}, we obtain the physical amplitude as
\begin{equation}
\label{amplitude_exact_1_appendix} {T}=R\tilde{T}_R, \qquad
\mbox{where}~\tilde{T}_R=\sum_{n=0}^\infty\tilde{T}_R^{(n)}\equiv
T^{\rm PWBA}+\sum_{n=1}^\infty\tilde{T}_R^{(n)}.
\end{equation}
Here $\tilde{T}_R^{(n)}=
\lim_{\Delta\rightarrow0}\tilde{T}_R^{(n)}(\Delta)$ is the
regularized on-shell Born term. The regularization function $R$ is
given by
\begin{equation}
\label{reg_function}R=\sum_{f'}\langle\delta({\bi r}_s)\delta({\bi
r}_e)\Phi_{f'}^{Z-1}|\Psi^{-}_{f}({\bi k}_s,{\bi k}_e)\rangle,
\end{equation}
where $\delta({\bi r})$ designates Dirac's delta function. The
function \eref{reg_function} has the following obvious property:
$R=1$ if $\tilde{\bfeta}=0$, where
$\tilde{\bfeta}=(\eta_s,\eta_e,\eta_{se})$.

Note that the above recipe for regularization is only one among an
infinite number of possible regularization procedures and, in
general, one obtains different regularization functions for
different regularization procedures. It means that while the
product $R\tilde{T}_R=T$~\eref{amplitude_exact_1_appendix} is an
algoristic quantity, the factors $R$ and $\tilde{T}_R$ are not,
i.e. $R\tilde{T}_R=\mathcal{R}\tilde{{T}}_\mathcal{R}$, where
$\mathcal{R}$ ($\mathcal{R}=1$ if $\tilde{\bfeta}=0$) and
$\tilde{{T}}_\mathcal{R}$ corresponds to an alternative
regularization procedure. For example, there is such
regularization procedure that yields $\mathcal{R}=1$ for any value
of $\tilde{\bfeta}$. To illustrate this statement, we expand $R$
in the Taylor series with respect to the components of
$\tilde{\bfeta}$ and notice that
$\tilde{T}_R^{(n)}\propto\tilde{\eta}^{n}$. We have
\begin{eqnarray}
\label{taylor_series} \fl R\tilde{T}_R=
\sum_{n=0}^{\infty}\frac{(\tilde{\bfeta}\cdot\bnabla_{\tilde{\eta}=0})^nR}{n!}\sum_{m=0}^\infty\tilde{T}_R^{(m)}
&=&\sum_{n=0}^\infty\sum_{m=0}^n\frac{(\tilde{\bfeta}\cdot\bnabla_{\tilde{\eta}=0})^{n-m}R}{(n-m)!}
\tilde{T}_R^{(m)}\nonumber\\\ms
&=&\sum_{n=0}^\infty\tilde{{T}}_{\mathcal{R}=1}^{(n)}=\tilde{{T}}_{\mathcal{R}=1},
\end{eqnarray}
where it is supposed that the operator
$$
\tilde{\bfeta}\cdot\bnabla_{\tilde{\eta}=0}=\eta_s\left(\frac{\partial}{\partial\eta_s}\right)_{\tilde{\eta}=0}
+\eta_e\left(\frac{\partial}{\partial\eta_e}\right)_{\tilde{\eta}=0}+
\eta_{se}\left(\frac{\partial}{\partial\eta_{se}}\right)_{\tilde{\eta}=0}
$$
acts only on the regularization function $R$, and
$(\tilde{\bfeta}\cdot\bnabla_{\tilde{\eta}=0})^{0}R=R|_{\tilde{\eta}=0}=1$.
As can be deduced,
$\tilde{{T}}_{\mathcal{R}=1}^{(n)}\propto\tilde{\eta}^{n}$ and
$\tilde{{T}}_{\mathcal{R}=1}^{(0)}=\tilde{T}_R^{(0)}\equiv T^{\rm
PWBA}$.

\section*{References}

\end{document}